%% file: main.tex
\RequirePackage[bookmarksnumbered,unicode]{hyperref}
\RequirePackage{hyperxmp}

\documentclass[conference]{IEEEconf}

\input epsf

\usepackage{todonotes}
\usepackage{csquotes}

\usepackage{caption}
\usepackage{subcaption}
\usepackage{amsmath}
\usepackage{booktabs}

\usepackage{cleveref}
\usepackage{lineno}
\usepackage{comment}

\RequirePackage{biblatex}

\newcommand{\wdym}{\textsc{wdym}}
\newcommand{\wdymFull}{\emph{What do you mean\dots?}}

\newcommand{\strong}[1]{\textbf{#1}}

\usepackage{graphicx}
\usepackage{multirow}
\usepackage{titlesec}
\usepackage{balance} 
\usepackage{stfloats}
\usepackage{xcolor}

\renewcommand\thesection{\arabic{section}} 

\renewcommand\thesubsectiondis{\thesection.\arabic{subsection}}

\renewcommand\thesubsubsectiondis{\thesubsectiondis.\arabic{subsubsection}}

\begin{document}

\include{data/questions_metadata.tex}

\title{What Does It Mean and Why Should I Bother?\\Motivating Students to Write Better Commit Messages}


\author{
    \IEEEauthorblockN{
        Gergő Balogh\IEEEauthorrefmark{1},
        Péter Seres\IEEEauthorrefmark{1},
        László Tóth\IEEEauthorrefmark{1},
        Attila Szatmári\IEEEauthorrefmark{1},
	Szakács Bence\IEEEauthorrefmark{1},
        Ádám Zoltán Végh\IEEEauthorrefmark{2}
    }
    \IEEEauthorblockA{
        \IEEEauthorrefmark{1}
        Department of Software Engineering,
        University of Szeged\\
        Árpád tér 2., 6720 Szeged, Hungary\\
        \{geryxyz, seresp, premissa, szatma, szakacs\}@inf.u-szeged.hu
    }
    \IEEEauthorblockA{
        \IEEEauthorrefmark{2}
        AENSys Ltd.\\
        Kálvária sgt. 24., 6722 Szeged, Hungary\\
        adam.vegh@aensys.hu
    }
}

\maketitle

\begin{abstract}
This paper reports on a locally motivated mixed-methods case study addressing a teaching-related suspicion held by software engineering instructors: that commit messages written by students frequently fail to serve their intended communicative role. To examine this suspicion empirically, we analyzed commit messages from student and industrial case-study projects using a partial replication of an established commit-message quality taxonomy. The results confirm that communication and quality issues occur recurrently in both contexts, substantiating the instructors’ initial concern. Motivated by this finding, we devised What Do You Mean? (WDYM), a lightweight, role-based educational game intended to surface and address commit-message communication breakdowns within the constraints of university coursework. Analysis of gameplay observations and participant surveys shows that WDYM is effective in raising awareness and fostering reflection on commit-message communication issues, although it provides limited evidence of immediate or sustained improvement in commit-message writing practices. Taken together, the study presents WDYM as a useful, though imperfect, context-bound intervention for addressing a locally observed instructional problem, without claiming general applicability beyond the studied setting.
\end{abstract}


\section{Motivation}
This paper reports on a mixed-methods case study conducted in the context of undergraduate software engineering courses at a European research university, based on the authors’ experience as course instructors.\footnote{To comply with double-blind review requirements, the institutional setting is described at an abstract level.} The study is grounded in a locally observed instructional predicament rather than a claim of general deficiency in software engineering education. In this setting, student-authored commit messages frequently fail to clearly communicate what was changed and why, revealing a recurring mismatch between the intended communicative role of commit messages and students’ actual practices.

From a teaching perspective, this observation raises a practical pedagogical challenge: how to convey the importance of commit messages to students who have not yet experienced the downstream costs of poor commit-message quality in long-lived or maintenance-intensive projects. While textbooks, lectures, and best-practice guidelines articulate principles for writing effective commit messages, such advice may remain abstract or slogan-like when detached from lived experience. We assume that experiential exposure can play a role in addressing this gap, while acknowledging that alternative instructional approaches may also be effective. Motivated by this challenge, we frame our work as a mixed-methods case study of the What Do You Mean? (WDYM) game, examining it as an educational intervention designed to make communication breakdowns around commit messages visible and discussable within the constraints of university coursework.

\section{Research Questions}
\label{sec:research-questions}

This paper investigates a locally observed teaching predicament and an educational intervention. The study is guided by two research questions that serve distinct but complementary purposes.

\strong{RQ1:} How and how frequently do communication and quality issues manifest in commit messages within student and industrial case-study projects, as observed through a replication of existing commit-message quality taxonomies?
This question aims to objectively assess a teaching-related suspicion derived from instructional experience, namely that commit-message communication problems occur frequently. By partially replicating established taxonomies and applying them symmetrically to student and industrial projects, RQ1 grounds this suspicion in empirical evidence. The industrial case-study data are used as a replication anchor and contextual reference, not as an intervention target.

\strong{RQ2:} To what extent does participation in the WDYM game surface awareness and sensitivity to commit-message quality and communication issues among participants?
This question examines WDYM as an educational intervention by analyzing participants’ awareness as it becomes visible through gameplay, survey responses, and reflections. All participants are treated as players engaging in the same interpretive task, while data are collected under different practical constraints for the two cohorts. For professionals, awareness is observed through pre-, post-, and follow-up surveys, allowing limited temporal comparison. For students, awareness is examined through post-game indicators and in-game observations only. Rather than reporting separate awareness outcomes per cohort, we present a unified analysis of participants, noting where evidence is longitudinal or cross-sectional depending on cohort-specific circumstances. The analysis focuses on perceptual and cognitive effects elicited by participation in the game, rather than behavioral change.

\section{Background and Related Works}
\label{sec:background-and-related-works}

Although our paper primarily focuses on a software engineering issue, it is important to note that our research has strong connections to theories in other disciplines. The nature of communication is a subject that falls within the domain of social and behavioral sciences like psychology and sociology. Also, pedagogy offers several theories about learning and teaching, and gamification has close ties to all these disciplines.

Gamification involves using game design elements in non-game contexts to create an interactive challenge that engages players at social, emotional, and cognitive levels,~\textcite{from-game-design-elements-to-gamefulness-defining-gamification,gamifying-software-engineering-tools-to-motivate-computer-science-students-to-start-and-finish-programming-assignments-earlier}. This technique is increasingly being adopted in educational settings to encourage student engagement,~\textcite{does-educational-gamification-improve-students--motivation--if-so--which-game-elements-work-best-}. In higher education, common game design elements include awarding points, badges, or levels based on achievement, implementing friendly competition via leaderboards, rankings, and duels, and engaging students with social interaction like role-playing and storytelling. Recent studies have shown that gamification can motivate students to commit more frequently,~\textcite{it-was-a-bit-of-a-race--gamification-of-version-control}.
Our research focuses on the quality of commit messages, with particular attention to their role as a communication artifact. In this study, we employ the What Do You Mean? (WDYM) game, which combines a round-based structure, explicit role assignment, and role-play, and player-driven scoring visualized through a leaderboard. Participants interpret commit messages individually and evaluate one another’s explanations through voting, forming a temporary scoring community without predefined rubrics or fixed quality criteria. This design allows interpretations and expectations of “good” communication to emerge from the group rather than being imposed a priori, and it enables the game to be readily adapted to different projects or participant groups. A lightweight narrative frames the activity and situates participants in a communication breakdown scenario, prompting reflection on how ambiguous or insufficient commit messages can lead to divergent understandings.

Interpreting and applying commit-message guidelines in everyday software engineering practice is challenging, particularly for novices. Many commonly recommended principles—such as clarity, relevance, and reader awareness—require contextual judgment that is difficult to formalize or apply without experience. As described by Polanyi, such skills exhibit the characteristics of tacit knowledge, as they are learned primarily through practice, observation, and immersion rather than explicit instruction,~\textcite{personal-knowledge-towards-a-post-critical-philosophy}. In the context of commit messages, developers refine these skills by encountering misunderstandings and observing how others interpret the same message. From a social learning perspective, novices acquire communicative competence by observing existing practices, attempting to apply them, and receiving feedback through misinterpretation or correction,~\textcite{social-learning-and-personality-development}. The WDYM game is designed to make this process explicit by exposing participants to situations where individually reasonable interpretations diverge, allowing them to observe misunderstandings as they emerge during play.

From a communication-theoretic perspective, commit messages can be understood as a specialized form of technical conversation that implicitly follows Grice’s cooperative principle,~\textcite{grices-cooperative-principle-meaning-and-rationality}. Effective commit messages tend to satisfy the maxims of quantity, quality, relevance, and manner, while violations—or floutings—of these maxims often introduce ambiguity and competing interpretations. WDYM operationalizes these breakdowns by requiring participants to interpret commit messages individually and evaluate the resulting explanations, thereby revealing how seemingly minor omissions or vagueness can lead to divergent understanding. Such ambiguity increases the cognitive effort required to comprehend changes, aligning with cognitive load theory, which posits that poorly structured information hampers task performance,~\textcite{cognitive-load-theory}. By making misinterpretations visible and discussable, WDYM provides experiential access to communication failures that would otherwise remain abstract.

\subsection{Quality of Commit Messages}

Writing proper commit messages is a tedious task. Therefore, several researchers provide automated methods for writing commit messages using machine learning tools and natural language processing methods, such as  \textcite{Mario2015}, \textcite{Liu2019}, \textcite{Xu2019}, and \textcite{Liu2022}. The main issue with these methods is that they lean on the existing, inappropriate commits.

Commit messages convey crucial information for the developers and code maintainers. Therefore, the commit messages should be clear, concise, and unambiguous. The knowledge sometimes also involves background knowledge that needs to be emphasized correctly. The SECI model, also known as the Nonaka-Takeuchi model, describes how background knowledge is converted into organizational knowledge, \textcite{Hoe2006}. Conveying background knowledge in short messages, like commit messages, is challenging, even if the authors provide informative descriptions. However, in many cases, the commit messages must offer more information for the code maintainers to catch the causes or goals behind the changes. In the article by \textcite{Reis2023}, the authors present their research results related to security changes, revealing that 56.7\% of the messages they found are poorly documented.

In order to support the developers, practitioners came up with a de facto standard called Conventional \textcite{conventional-commits}. The term relates to conventions for writing clear and consistent commit messages that can be used to automate tools and communicate changes. \textcite{jung-2021-commitbert} applied the idea of these conventions to create a 345K dataset consisting of code modification and commit messages in six programming languages (Python, PHP, Go, Java, JavaScript, and Ruby). Besides the dataset, the author created a model called commitBERT, which is capable of fabricating commit messages automatically.

However, a relatively small number of studies provide comprehensive descriptions of the quality of the commit messages. The paper by \textcite{what-makes-a-good-commit-message} is one of the research studies that defines a taxonomy that can be used for measuring the quality attributes of commit messages. Next to the taxonomy definition, the study presents the result of investigating the commit messages' qualities in various Java open-source projects. The authors examined 1,600 messages and found that circa 44\% of these messages could be improved. The paper written by \textcite{Li2023} provides semantic measures for the quality of commit messages. The authors developed a classifier to automatically measure quality, outperforming the state-of-the-art machine learning classifiers by 12 percentage point in the F1 score. 

The OSS projects provide many examples for studying the commit messages. \textcite{Chahal2018} explored the impact of community dynamics on the syntactic quality of commit messages of an OSS project. The empirical study of the 7 OSS projects shows that only a small number of active contributors lead to high syntactic quality contributions. \textcite{Motta2018} focused on the architectural information conveyed by the messages. It was found that a small number (232 out of 42,117) contained project decisions and architectural aspects that motivated changes in the project.

\textcite{improving_the_quality_of_commit_messages_in_students_projects} focused on improving the commit message writing skills of the students by modifying the GitHub desktop to present the crucial \emph{\enquote{what}} and \emph{\enquote{why}} parts to the students. The authors found that highlighting these essential parts enhances the quality of the messages.

\section{Experimental Design}
\label{sec:experimental-design}

\begin{figure}[h]
	\centering
	\includegraphics[width=\linewidth]{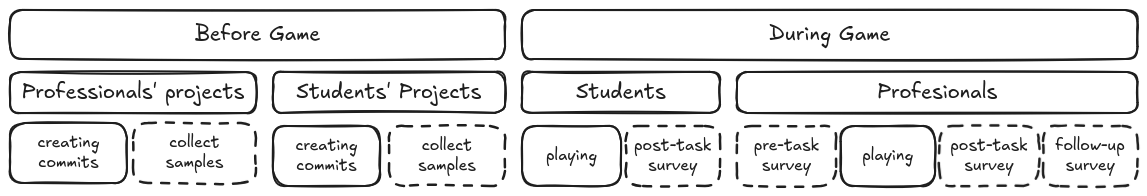}
	\caption{Overview of the Research's Phases}
	\label{fig:research_overview}
\end{figure}

Our research consisted of several phases to collect the necessary qualitative and quantitative information and answer the research questions (\Cref{fig:research_overview}). We have collected commits from the participating students' and professionals' previous projects to evaluate how frequently and in what way commit messages fail. By doing so, we partially replicated the experiment conducted by \textcite{what-makes-a-good-commit-message}. We used the labeling system created by \textcite{what-makes-a-good-commit-message} to label the non-merge commits in the samples and those utilized during the games. The \enquote{During Game} phases involved two game sessions with students and professionals, which were preceded and followed by surveys (pre, post, and follow-up).
We randomly selected 104 student commit messages and 33 professional commit messages for manual inspection, reflecting the available expert annotation capacity. Merge commits were excluded, as they are not suitable for evaluation under the adopted labeling scheme. This inspection was not intended to yield statistically significant or generalizable results; rather, it served as an exploratory measurement to corroborate prior observations by course instructors regarding commit message quality within the local academic and professional context.

We used a qualitative analysis method to gather the opinions of five software engineering experts. Each expert independently chose the labels for each commit message. Their votes were counted, and the ratio of each label per item was calculated.

\begin{figure}
	\centering
	\includegraphics[width=\linewidth]{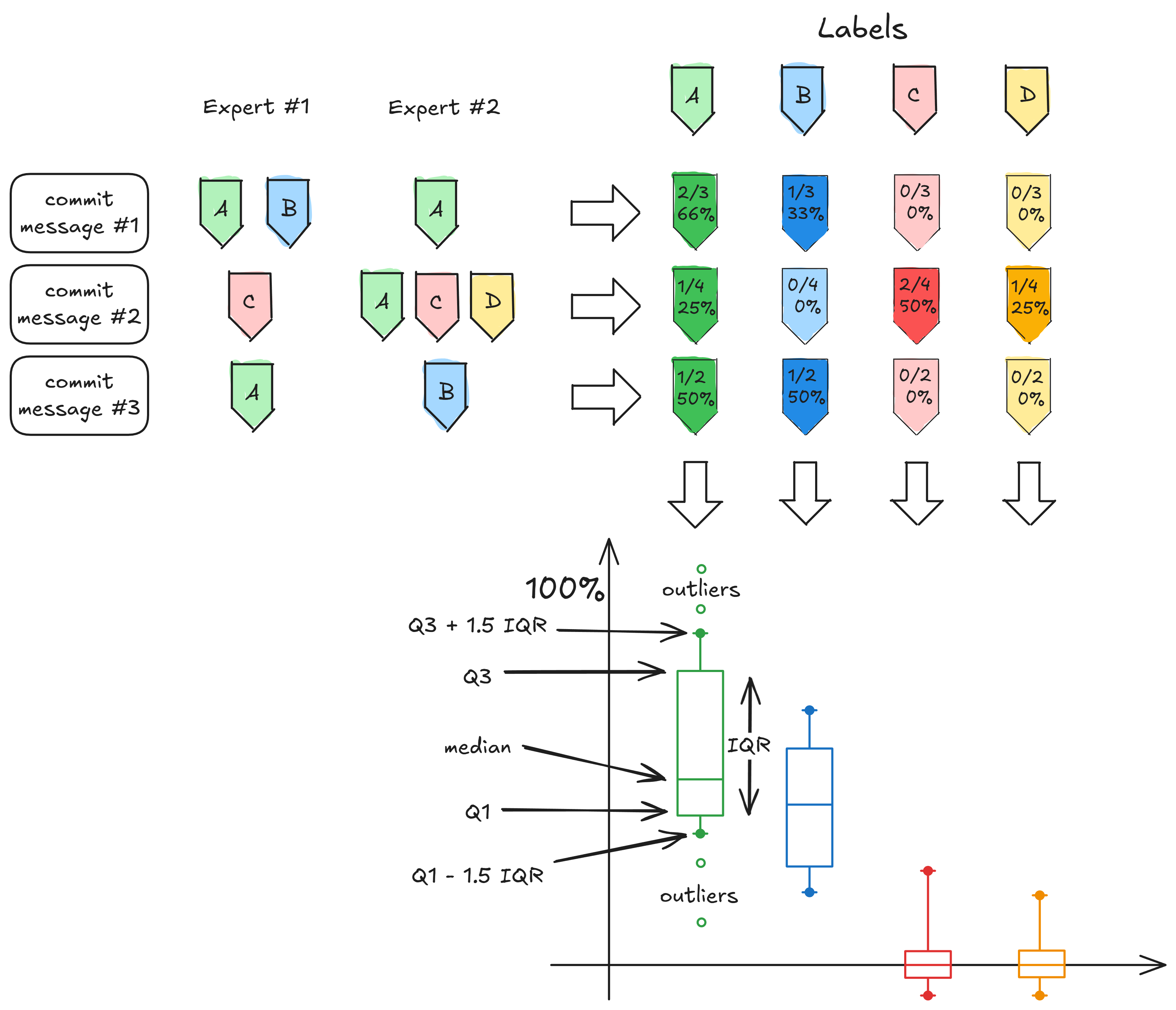}
	\caption{Multi-label Majority Voting Algorithm}
	\label{fig:labeling}
\end{figure}

For example, consider \Cref{fig:labeling}. In this case, two experts gave three votes for commit message \#1. They gave two votes for labels A, one for label B, and none for labels C and D. Therefore, the aggregated percentages for commit message \#1 are approximately 66\%, 33\%, 0\%, and 0\% for labels A, B, C, and D.
We present the distribution of percentages per label on boxplot charts (e.g., \Cref{fig:commit-labels}). The legend for these charts is located in the lower part of \Cref{fig:labeling}. Please note that the total percentage per label can exceed 100\% as a result of multiple labelling.

\subsection{The \wdymFull{} Game}
\label{sec:the-wdym-game}
The \wdym{} game is a paper-and-pen role-playing game that simulates a story. You play the role of an inexperienced software developer who has identified a commit that will likely be useful in your work. However, you do not entirely understand what the commit is all about. You ask your more experienced colleagues to explain it, but everyone is too busy to provide a detailed explanation. Instead, they give you brief text responses via Teams, Skype, or Discord. To simulate a pessimistic scenario, the game's logic assumes that experienced colleagues only rely on the commit message and not the content of the commit; they are extremely busy, so to speak. After that, you need to decide whether their answers are helpful and provide enough clarity for you to use the commit effectively.

\begin{figure}[h]
	\centering
	\includegraphics[width=\linewidth]{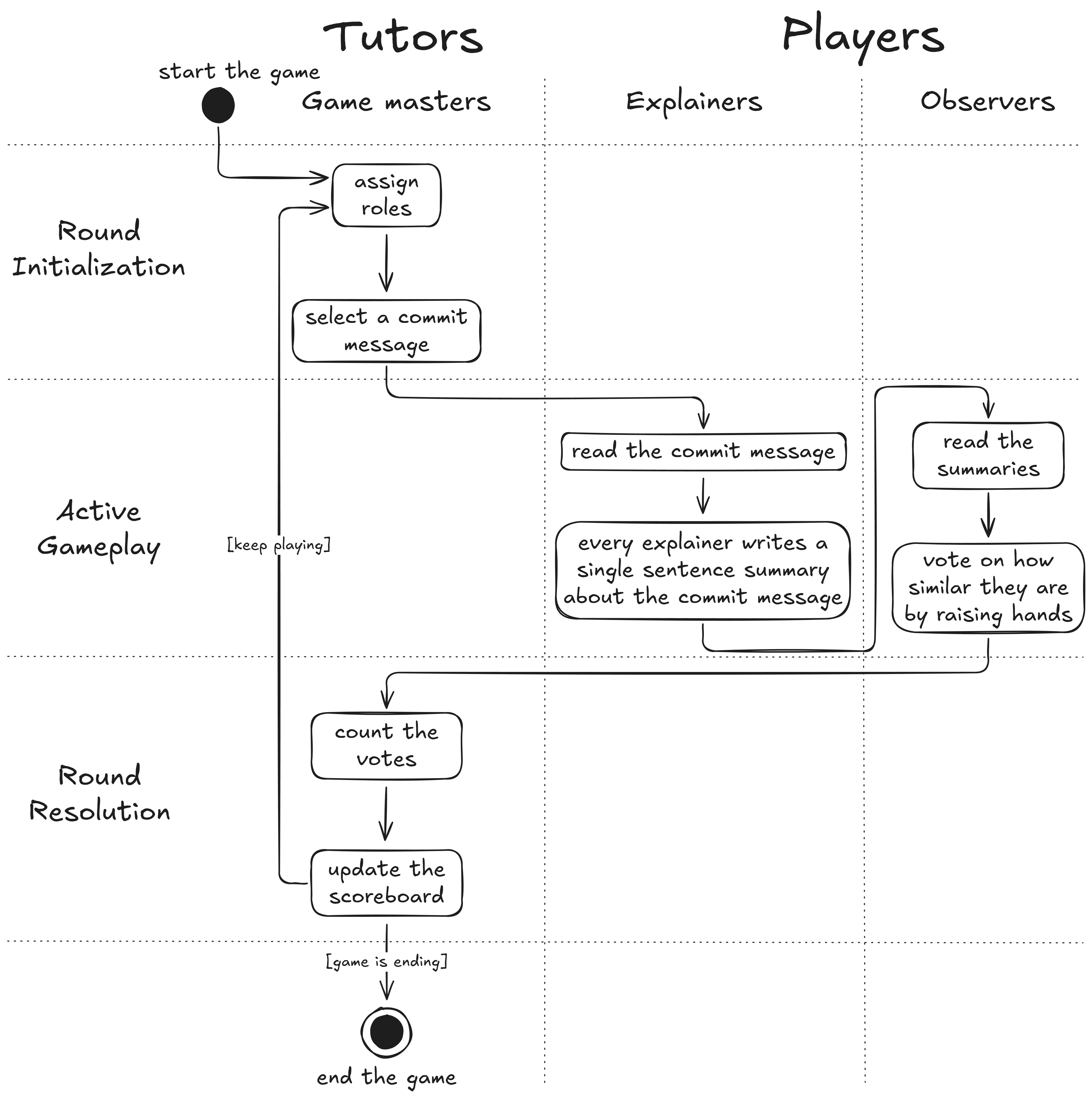}
	\caption{The \wdymFull{} game}
	\label{fig:wdym-game}
\end{figure}

In this paper, we use role names that differ slightly from those used in the acted game narrative to avoid ambiguity in the methodological description. The game master coordinates each round by assigning roles and selecting a commit message, which is read aloud. Explainers (referred to as colleagues or seniors in the story) independently interpret the commit message and produce brief written explanations or summaries. Observers (juniors in the story) then review these explanations, interpret their meaning, and vote on whether the explanations are consistent. This terminology is used consistently throughout the paper to clearly distinguish functional roles in the game from their narrative counterparts.

The game consists of several rounds, each with the same steps shown in \Cref{fig:wdym-game}. There are several free parameters of the game, which were set based on our prior experience. These parameters are mainly tied to the scoring system, number of players, and selection of the commit. Each round, the game masters assign the roles of the players and choose a commit from the subject system. Next, the explainers (mentors in the story) read only the commit message and write a brief summary about it. The observers (mentees in the story) read all the summaries. The summaries could be read out loud, passed from observer to observer, or even projected onto a whiteboard. The only crucial restriction is that explainers are not allowed to change their summaries while sharing them. The observers then vote on the answers for consistency. They can vote for the summaries being similar (yes vote) or different (no vote) in meaning by raising their hands. Similar summaries suggest an easily interpretable commit message, while summaries different in meaning tell us, that the commit message failed to achieve its goals: communication while observing Grice maxims. The game masters then count the votes and update the scores before starting the next round.

\begin{figure}[h]
	\centering
	\includegraphics[width=\linewidth]{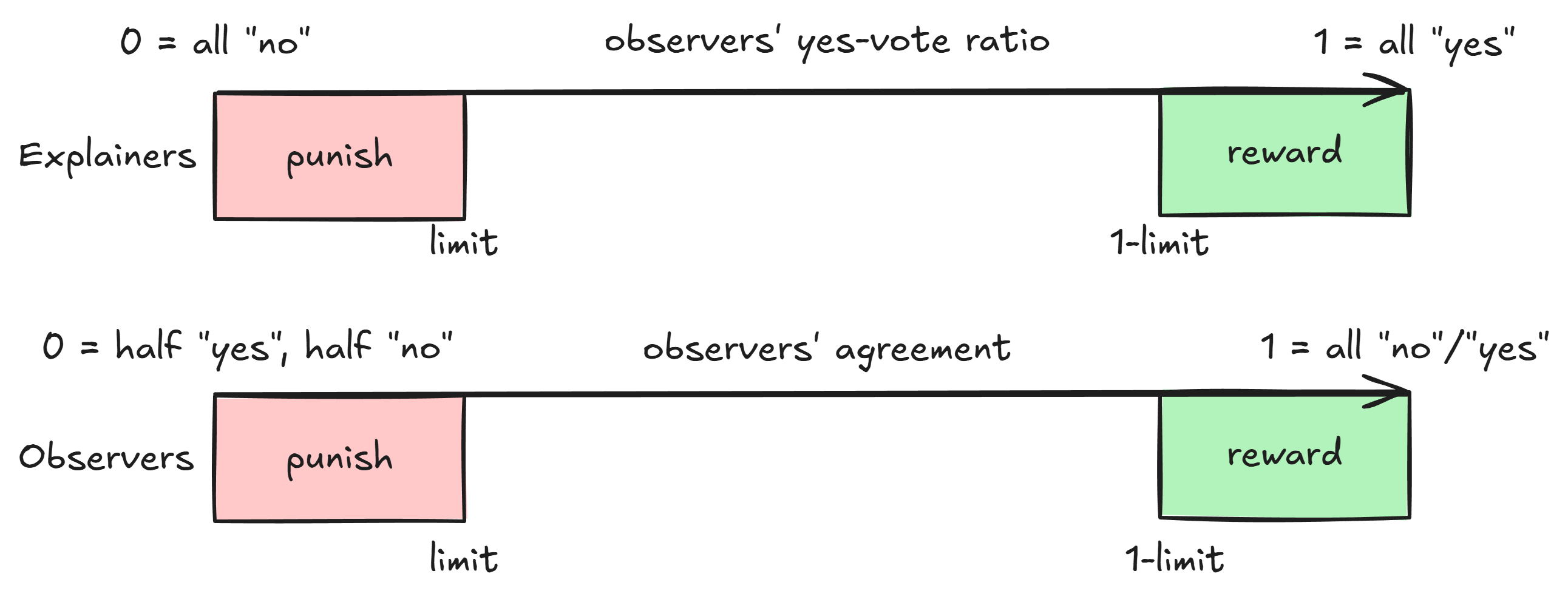}
	\caption{The \wdymFull{} game's scoring logic}
	\label{fig:wdym-game-scoring}
\end{figure}

The scoring mechanism, illustrated in \Cref{fig:wdym-game-scoring}, is designed to reward clear communication by explainers and careful, consensus-driven judgment by observers. Explainers gain or lose points based on the observers’ yes-vote ratio: a high proportion of “yes” votes indicates that independently written summaries converged in meaning and is therefore rewarded, while a ratio below a predefined threshold is penalized. Observers are scored differently, based on their internal level of agreement rather than the direction of their votes. Agreement is calculated as the distance from a split decision, where unanimous or near-unanimous voting—whether “yes” or “no”—yields high agreement and is rewarded, while evenly divided votes are penalized. This design encourages observers to take the task of judging semantic similarity seriously, as careful interpretation should lead to shared conclusions, and it discourages unreflective strategies such as indiscriminately voting “no” in every round. Rewards and penalties are symmetric for explainers (±5 points) and reduced for observers (±3 points), allowing observers to remain competitive while preserving the explainers’ primary responsibility for communicative clarity. The reward and punishment thresholds for both explainers and observers were defined by lower and upper limits set at 33\%.

To examine the effects of the WDYM game, we administered a set of surveys designed to capture participants’ perceptions, interpretations, and reflections related to commit-message communication. Survey administration followed a phased structure consisting of post-game surveys for all participants, with additional pre-game and follow-up surveys conducted for professional participants where practical constraints allowed.

\noindent\paragraph{Pre-game.}
A pre-game survey was administered only to professional participants. Its purpose was to establish a baseline of prior experience with commit messages and perceived communication issues in everyday development work. A comparable pre-game survey was not conducted for students, as their experience with commit writing is typically limited, heterogeneous, and closely tied to course assignments, making a stable and interpretable baseline difficult to obtain within the available instructional timeframe.

\noindent\paragraph{Post-game.}
Immediately after the game sessions, all participants completed a post-game survey. This survey focused on awareness and recognition of commit-message communication issues, interpretation of ambiguous messages, and subjective experience of the game. For students, this post-game survey constitutes the primary source of evidence, capturing how awareness is surfaced through participation in the game rather than measuring change over time.

\noindent\paragraph{Follow-up.}
A follow-up survey was administered one week later to professional participants only. This survey aimed to observe short-term persistence of awareness in an industrial context where participants regularly engage in version control activities. A follow-up survey was not conducted for students, as they typically complete commit-related coursework within a limited time window and do not consistently engage in development activities after the intervention.

\subsection{Participants}

For our experiment, we recruited two groups of participants based on their level of programming knowledge and professional experience. Professional participants were recruited through invitations sent to company management, after which developers could freely sign up without filtering or selection. Student participants were recruited from undergraduate courses involving version control and team-based software development. Students were not filtered based on prior performance or experience.

In total, 157 participants took part in the study. Of these, 143 were students, primarily enrolled in the final year of their undergraduate program, and 14 were professional developers. Among the professionals, six were classified as junior developers, one as mid-level, and seven as senior developers. Participants in both groups primarily worked with Java or JavaScript, and most professional participants identified as back-end developers.

While students constituted the majority of participants and contributed the primary commit-message data used to address RQ1, survey participation differed substantially between the two cohorts. Due to curriculum constraints and ethical considerations, student participation in surveys was voluntary and could not be enforced without introducing additional workload or instructor-induced pressure. As a result, only 11 students completed the post-game survey. In contrast, all 14 professional participants completed the applicable survey phases. Because professional survey responses were collected under less constrained conditions and with higher response completeness, they provide a more reliable basis for interpreting awareness-related effects of the game. Student survey responses are therefore treated as supplementary and contextual, while student participation in gameplay and commit-message analysis remains central to the study.

\subsection{Subject Systems}
During our experiment, we collected 19 open- and closed-source projects.
Subject systems were selected so that at least some of the participants have prior knowledge about them, preferably having some prior contribution.
We collected this information directly from the participants during the sign-up phase.
In the case of students, we also utilized projects that are part of their course assignments.
These subject systems were used to showcase the commit message problem and evaluate our game.

Most of these programs are written in Java, and most have a fair amount of commits, making them suitable for our experiment. The projects varied significantly in size, with the smallest being just 0.6 thousand Lines of Code (kLOC) and 50 commits and the largest reaching up to 13,869 kLOC and 84,707 commits. On average, projects comprised approximately 989 kLOC and 5,877 commits. However, the median values, which better represent the typical project size due to the wide range, stood at 28 kLOC and 564 commits. The list of the program names and other details can be found in the online appendix. Some of these programs were not open-source, therefore, we had to leave out some details from the list.

\section{Replication Package}

A replication package\footnote{The supplemental material is provided in the \textbf{\textit{online\_appendix.zip}} \url{https://doi.org/10.5281/zenodo.18787232}.} has been created and made available in the online appendix to allow interested readers to learn more about the results and the surveys in general. The package contains two directories. One is for the commit labeling process, which contains the diagramm shown in \Cref{fig:commit-labels}. Note that we could not include the commits themselves as they were from closed-sourced proprietary projects. The other directory contains two subdirectories, one for the survey forms in Hungarian and the other for the results of the surveys, and a file with the translations of the survey questions and survey metadata.

\section{Results}
\label{sec:results}

\subsection{Commit-Message Communication Issues}

\begin{figure*}
	\centering
	\includegraphics[width=\textwidth]{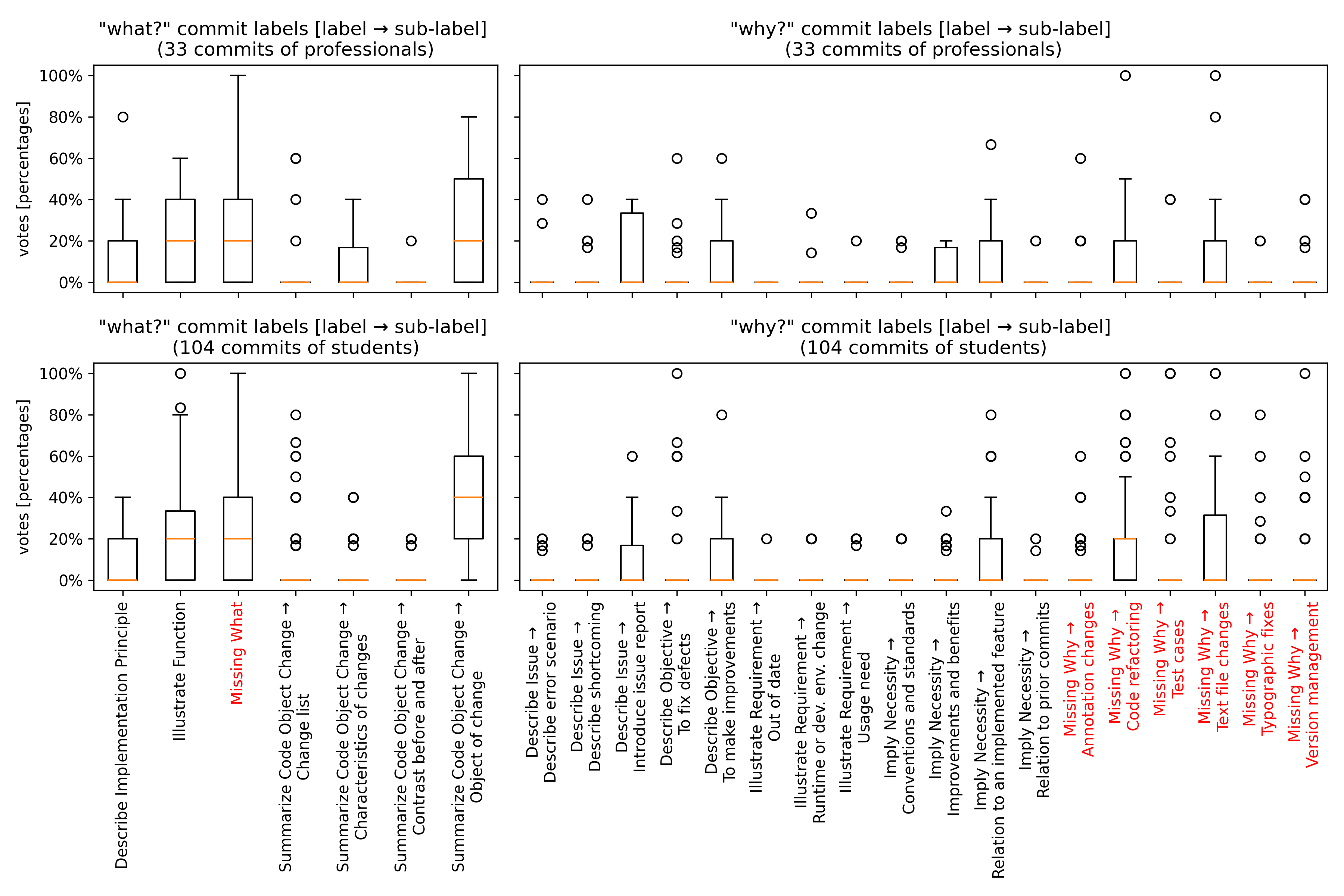}
	\caption{Commit messages' labels of students and professionals}
	\label{fig:commit-labels}
\end{figure*}

Based on the distribution shown in \Cref{fig:commit-labels}, RQ1 empirically validates prior observations by course instructors that commit-message communication and quality issues are not sporadic but recurrent in both student and industrial projects. Across both cohorts, multiple labels associated with missing or insufficient information exhibit non-zero medians and broad upper quartiles, indicating that such issues occur across a substantial fraction of commits rather than appearing as isolated outliers. While the overall label distributions diverge between students and professionals—suggesting differences in how communication breakdowns manifest—the leading labels overlap in important ways. In particular, \enquote{Missing What} emerges as the second most frequent label in both cohorts, showing that commits often fail to clearly state what was changed even when some description is present. Labels related to \enquote{missing why} information display a more heterogeneous pattern, yet several of them still appear among the dominant categories for both groups. Taken together, these distributional patterns demonstrate that communication problems manifest through systematic omissions of essential information and that they do so with sufficient regularity to corroborate the initial teaching-related suspicion within both academic and professional contexts.

\textbf{We answer RQ1 by showing that communication and quality issues manifest frequently in (locally sampled) commit messages in both student and industrial case-study projects, with comparable categories of issues recurring across contexts, which validates prior observations by course instructors.}

\subsection{Awareness of Communication Issues after Gameplay}

In RQ2, awareness of communication issues is operationalized through participants’ self-reported reflections captured in post-game (and, where applicable, pre- and follow-up) questionnaires (\Cref{tab:rq2-questions}), which were designed to surface perceptual and cognitive effects rather than behavioral change. The surveys contained a broader set of questions than those reported here in order to efficiently reuse the collected data for multiple research purposes; questions not directly relevant to RQ2 are therefore intentionally omitted from this subsection and reserved for future analyses. All questionnaires are provided in the online appendix in their original Hungarian language for transparency and replication. For the purpose of answering RQ2, we group the relevant items into three conceptual categories: (1) direct questions about the perceived effect of the game, probing whether and how participants believe the game influenced their thinking or future practices (e.g., whether they expect to write commit messages differently after gameplay); (2) direct questions about past experience, eliciting participants’ prior attitudes and encounters with similar communication situations; and (3) indirect questions focusing on noticing and recognizing the situation, capturing surprise, recognition, or perceived prevalence of such communication breakdowns as revealed during gameplay. This grouping provides the analytical structure for the following subsection, where we discuss the participants’ responses in detail.

In this study, awareness is understood as a surfaced sensitivity to commit-message communication issues, made observable through participants’ self-assessments of their past practices, their recognition of similar situations, and their stated intentions regarding future commit-message writing. Impact is interpreted narrowly as the perceived relevance of the game experience to participants’ professional thinking and reflection, rather than as evidence of sustained behavioral change.

\begin{table*}[h]
	\small
	\centering
	\caption{Survey questions used to operationalize RQ2}
	\label{tab:rq2-questions}
	\begin{tabular}{@{}ll@{}}
		\toprule
		\textbf{Question} & \textbf{Type} \\
		\midrule
		\multicolumn{2}{c}{\textbf{Direct questions about the perceived effect of the game}} \\
		Do you think the experiences gained during the game will have an impact on your non-professional life? & Open-ended \\
		To what extent did the game experience impact to your professional development in software development? & Predefined (scale) \\
		Justify how you assessed the impact of the game on your professional development. & Open-ended \\
		Will you write commit messages differently after the game? & Yes/No \\
		Why did you make these decisions regarding future commit messages? & Open-ended \\
		\midrule
		\multicolumn{2}{c}{\textbf{Direct questions about past experience}} \\
		What is your opinion on the clarity of your own commit messages? & Predefined (scale) \\
		Please explain why you made the decisions you did regarding your commit messages. & Open-ended \\
		Have you been in a similar situation before? If yes, in what role? & Predefined (game's roles) \\
		\midrule
		\multicolumn{2}{c}{\textbf{Indirect questions on noticing the situation}} \\
		Did the responses of colleagues and authors to commit messages surprise you? & Predefined (scale) \\
		How often do software developers find themselves in similar situations? (reflecting on the game's story) & Predefined (time-scale) \\
		\bottomrule
	\end{tabular}
\end{table*}

\subsubsection{Direct Questions about the Perceived Effect of the Game}

\begin{figure}[h]
	\centering
	\includegraphics[width=\linewidth]{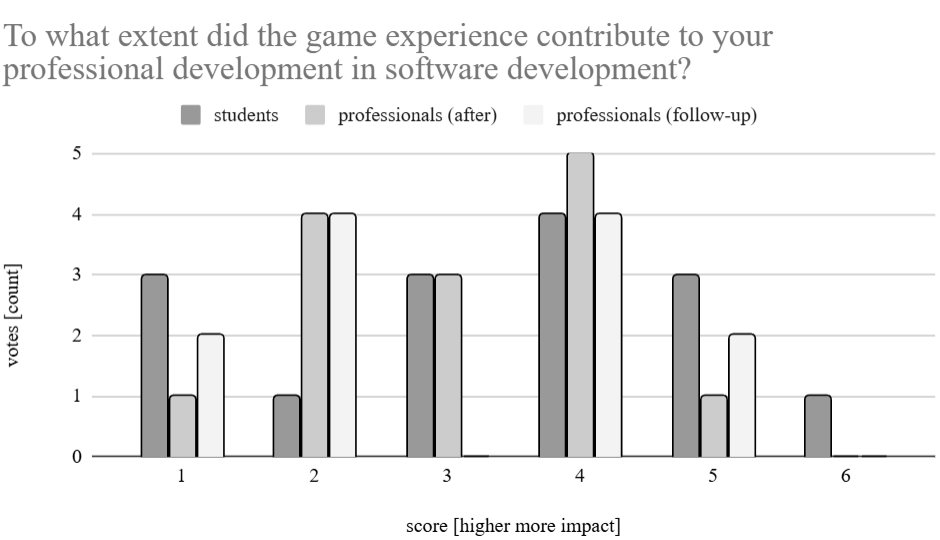}
	\caption{Self-reported impact about professional development}
	\label{fig:professional-development}
\end{figure}

The direct questions targeting the perceived effect of the game indicate a moderate but consistent reflective impact across participants. When asked to assess the game’s contribution to their professional development (Figure \ref{fig:professional-development}), both students and professionals clustered their responses around the mid-range of the scale (typically 3–4 out of 6), suggesting that the game was perceived as meaningful but not transformative. This pattern was stable across cohorts, and follow-up responses from professionals show a slight attenuation over time, indicating that the perceived impact diminishes but does not disappear shortly after gameplay. Qualitative justifications reveal that participants primarily valued the game for broadening their perspective on how others interpret commit messages and for exposing different styles and expectations, rather than for teaching concrete commit-writing techniques; several respondents explicitly noted that if relevant information is absent from the commit message, effective communication becomes impossible regardless of intent.

\begin{figure*}[h]
	\centering
	\includegraphics[width=.7\linewidth]{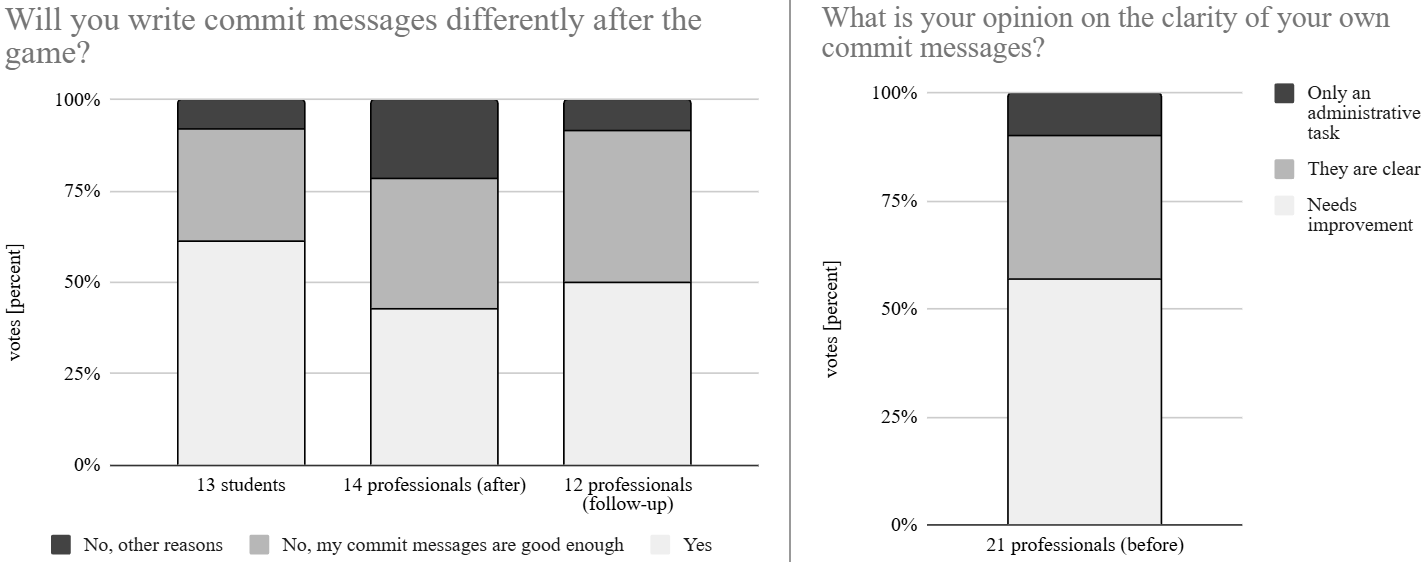}
	\caption{Self-reported quality of past commit messages and intent to change them}
	\label{fig:past-commit-future-commit}
\end{figure*}

Intent-related questions further support this interpretation of awareness as reflective rather than behavioral. Approximately half of the participants reported an intention to write commit messages differently after the game (Figure \ref{fig:past-commit-future-commit}), while fewer than a quarter reported no intention to change. Importantly, the latter group did not uniformly justify their stance by claiming their commit messages were already of high quality; instead, responses ranged from confidence in current practices to skepticism about the feasibility of capturing complex changes in short messages. Participants who expressed an intention to change most frequently attributed this to in-game examples that prompted reevaluation of their own commit-message quality—either by revealing shortcomings or by confirming previously implicit good practices—often accompanied by a general sentiment that there is always room for improvement. Together, these results suggest that the game primarily surfaces reflection and sensitivity toward commit-message communication, rather than inducing immediate or uniform plans for behavioral change.

\subsubsection{Direct Questions about Past Experience}

The direct questions about past experience capture participants’ self-assessments of their prior commit-message practices and their familiarity with situations similar to those enacted in the game. At an aggregate level, the distribution of responses regarding the perceived clarity of past commit messages resembles the distribution of stated future intent reported in the previous subsection (Figure \ref{fig:past-commit-future-commit}). However, because individual participants were not tracked across instruments, these results do not allow any inference about within-participant change or alignment; the observed similarity therefore reflects cohort-level patterns only and cannot be interpreted as evidence of opinion shift or stability at the individual level.

Participants’ explanations for their assessments reveal recurring, context-driven rationales rather than a single dominant deficiency. Commonly cited reasons for perceived lack of clarity include time pressure, solo development contexts, the perceived low impact of certain changes (e.g., typo fixes or refactorings), and the difficulty of adequately summarizing complex modifications within short commit messages. Several participants also expressed the view that commit messages, by their nature, cannot fully capture the meaning of code changes, while others emphasized that overly long commit messages can themselves hinder understanding. These responses suggest that participants’ evaluations are grounded in practical trade-offs rather than simple neglect or lack of awareness.

\begin{figure*}[h]
	\centering
	\includegraphics[width=.7\linewidth]{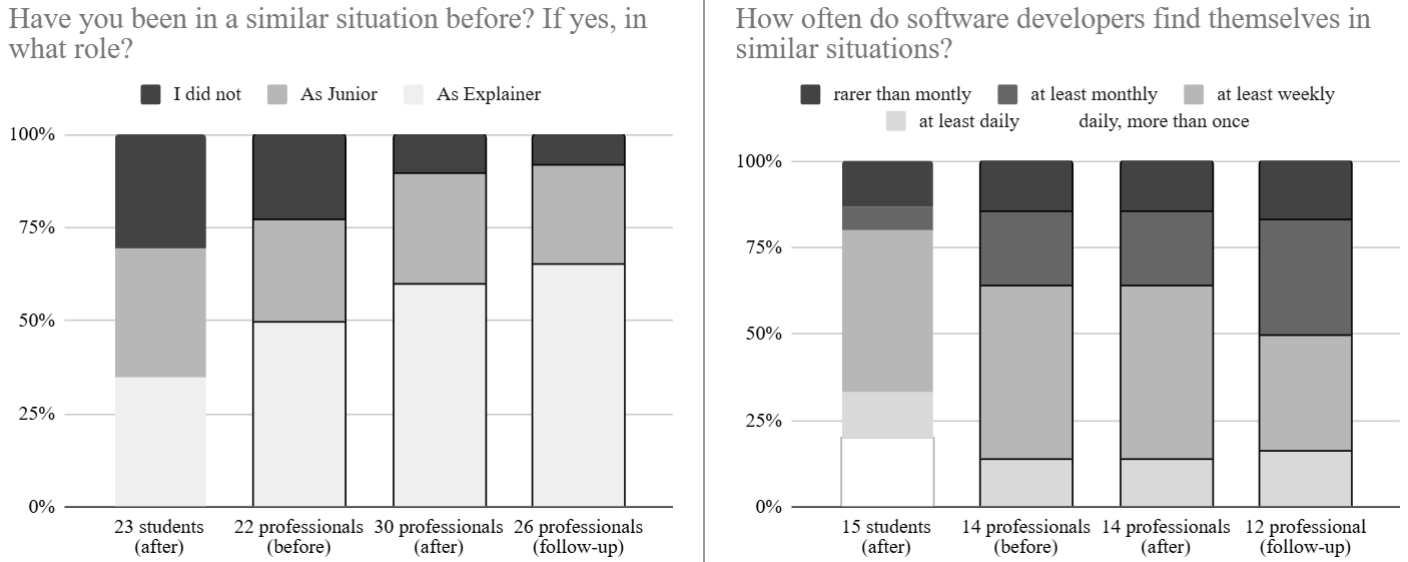}
	\caption{Estimated frequency of game situation about self and others}
	\label{fig:game-situation}
\end{figure*}

When asked about prior exposure to situations similar to the game’s role-play scenario, participants reported increased recognition of such situations, particularly when reflecting on their own past experiences (Figure \ref{fig:game-situation}). As expected, students reported fewer prior encounters than professionals; however, the difference between the two cohorts was smaller than initially assumed. This indicates that while professional experience increases exposure to commit-message communication breakdowns, even less experienced participants can relate the game scenario to their own development contexts, supporting the relevance of the enacted situation across experience levels.

\subsubsection{Indirect Questions on Noticing the Situation}

The indirect questions address whether gameplay made participants notice or bring into focus commit-message communication issues, even in the absence of strong emotional reactions. When asked whether the explanations provided by colleagues and authors during the game were surprising, both students and professionals reported low levels of surprise. Only a minority of participants indicated being surprised by the responses (students: 4 out of 11; professionals: 1 out of 13), suggesting that the observed misunderstandings and divergent interpretations were largely consistent with participants’ existing expectations rather than perceived as novel phenomena.

Importantly, the low reported surprise does not imply that the game was ineffective in surfacing the issue. Several responses indicate that, although the situations themselves were not unexpected, the game format made the communication breakdowns more explicit and easier to attend to by juxtaposing multiple interpretations side by side. In this sense, awareness is expressed not through novelty but through focused recognition of a familiar problem.

This interpretation is further supported by responses concerning the perceived frequency of similar situations among software developers in general. Self-reported estimates of how often others encounter such situations remain largely stable after gameplay, indicating that participants did not revise their general beliefs about prevalence. At the same time, this stability contrasts with the increased recognition of having personally encountered similar situations discussed in the previous subsection, suggesting a growing differentiation between abstract beliefs about “typical developers” and reflective awareness of one’s own experience. These responses suggest that the game brings a familiar issue into sharper focus rather than altering participants’ beliefs about how common such situations are.\bigskip

\textbf{We answer RQ2 by showing that participation in the WDYM game surfaces awareness of commit-message communication issues through increased recognition and reflection, while yielding limited evidence of strong perceived impact or immediate behavioral change.}

\section{Threats to Validity}
\label{sec:limitations}

As with any empirical study, this work is subject to several threats that may affect the validity of its findings. In this paper, we consider potential issues related to locality, commit message labeling, selection of participants, survey design, and survey traceability.

Our survey instruments were not validated previously, so, although aligned with the research questions, they may not capture the multifaceted construct of "awareness" as intended. In our study, we view awareness as the interest in improving, looking for, and identifying a specific problem. 
The self-reported measures may introduce the risk of social desirability bias, interpretation bias, and the dual role of instructors as researchers could further influence the behavior of the student participants.
In order to collect as much valuable information as possible, we mostly asked questions to which we expected an open response.

Although multiple experts independently labeled the commits and discussed disagreements, and open-ended survey responses were categorized collaboratively by the researchers, we did not compute formal inter-rater agreement statistics for either process. Given the exploratory and locally bounded case-study design, both the commit labeling and the qualitative survey coding were intended to provide structured triangulation and thematic interpretation rather than statistically generalizable or fully reproducible classification results.
The commit labeling process, when dealing with ambiguous or overlapping commits, and the categorization of open-ended questions both required subjective judgement. All of the controversial cases in both commit labeling and survey answer categorization were discussed in reconciliation meetings.

In this local case-study we did not filter any of the participants. The students and the professionals all volunteered, meaning that the participants might be more reflective than their peers. In the case of the students, we could not make the surveys mandatory as we would run into ethical problems, and could also skew the results of the survey or introduce more desirability bias. Furthermore, this voluntary participation means we could not trace the survey answers throughout the questions for the students. Tracing answers could have given us a broader, better perspective on the awareness amongst students.
Differences in project maturity, scale, and expertise may confound the comparison of label frequencies. It must also be highlighted that in the case of the student groups, we selected commits for labeling from projects that were created during university courses. Since these projects are not always scored with the quality of the commit messages in mind, students might have disregarded the importance of writing good commit messages.

Considering that we did not want to measure improvement, the addition of a control group could have given us another viewpoint. How much the \wdym{} game helped raise awareness compared to talking about the problem is up for debate. Gamification itself could have also introduced some temporal elevation of awareness independent of the content of the game.

It is important to note that the results of this study are not general; this is a non-representative case-study of Hungarian university students and professionals. While the findings are context-bound, the study design and game mechanics are described in sufficient detail to support replication in other contexts.
The translation of survey instruments (reported in Hungarian) may hinder the interpretation for international readers and future replications.
\section{Discussion}
\label{sec:lessons-learned}

Based on the information gathered, we discovered that both the students and the professionals write commit messages that fail to communicate their intentions. These lousy commit messages could lead to communication issues, which could hinder productivity or create additional workload for the author of these commit messages. Although both groups are aware of this issue, their attitudes, mindsets, and motivations toward addressing it differ. The self-reported acknowledgment and observation of the simulated scenario showed awareness after the game, possibly because the scenario persisted in the participants' minds. However, the effectiveness of our inaugural edition of the \wdym{} game varies depending on the players' preferences. As shown before, participants reported that playing the game was a meaningful event, but it did not transform their views on the topic entirely. The data let us assume that the game's current edition is beneficial to anyone who is interested in exploring the commit messages and the mindset of their peers, but we cannot state that it will boost motivation and awareness in general.

Experts reported that labeling students' commit messages was challenging, since their commits did not match the labels created for professionals' work. At times, experts could not determine if the commit message contained information relevant to the labeling system. Deciphering such messages required more time and effort, and also further emphasized how communication issues create additional workload. Understanding these commit messages depends heavily on the available resources. Knowing the commit author's skills makes the understanding and labeling more efficient.

The research results presented in this study reinforce our assumptions that students and professionals do not pay enough attention crafting quality commit messages, and students require additional context (or first-hand experience) on how these poorly written commit messages could affect their development experience. We believe these issues are connected to the hard problem of simulating real software development in an educational setting.

\section{Future Work}

We aim to pursue multiple directions in our future research related to this theme. Firstly, we will continue enhancing our game prototype to aid anyone who is interested in improving their understanding and writing commit messages. An online web application is already in development. Secondly, we want to expand beyond our local environment and measure how these problems manifest in other groups. We might want to introduce new methodologies such as LLM-assisted commit labeling, an improved and traceable survey system, and the enhancement of the player experience.

Our long term goal is to provide practical solutions for tutors and a great learning opportunity for students, juniors, or any developer who wants to improve on these problems.

\section{Acknowledgment}

We would like to thank Péter Seres for implementing the software support and evaluation scripts used in the experiment, and for his valuable assistance in organizing and executing the study.

\printbibliography


\end{document}

%% file: data/questions_metadata.tex
\newcommand{\qMakeStoryClear}{\enquote{Make Story Clear} [open]}
\newcommand{\questionMakeStoryClear}{\enquote{How could this story be told more vividly to make it more understandable to even more people?} [open]}
\newcommand{\qYearsInTech}{\enquote{Years in Tech?} [open]}
\newcommand{\questionYearsInTech}{\enquote{How many years have you been working as a software developer or programmer?} [open]}
\newcommand{\qEvaluateGameSImpact}{\enquote{Evaluate Game's Impact} [open]}
\newcommand{\questionEvaluateGameSImpact}{\enquote{Justify how you assessed the impact of the game on your professional development.} [open]}
\newcommand{\qPriorKnowledgeOfTheGame}{\enquote{Prior Knowledge of the Game?}}
\newcommand{\questionPriorKnowledgeOfTheGame}{\enquote{Were you familiar with the game before?}}
\newcommand{\qStatementsAboutYourJob}{\enquote{Statements about Your Job}}
\newcommand{\questionStatementsAboutYourJob}{\enquote{Mark the statements that apply to you.}}
\newcommand{\qPlayedBefore}{\enquote{Played Before?}}
\newcommand{\questionPlayedBefore}{\enquote{Have you played this game before?}}
\newcommand{\qJustifyCommitMessageChoices}{\enquote{Justify Commit Message Choices} [open]}
\newcommand{\questionJustifyCommitMessageChoices}{\enquote{Please explain why you made the decisions you did regarding your commit messages.} [open]}
\newcommand{\qUnderstoodStory}{\enquote{Understood Story?}}
\newcommand{\questionUnderstoodStory}{\enquote{Did you understand the story?}}
\newcommand{\qWhichCourseSection}{\enquote{Which Course Section?}}
\newcommand{\questionWhichCourseSection}{\enquote{Which course sections did you play in?}}
\newcommand{\qMostMemorablePart}{\enquote{Most Memorable Part?} [open]}
\newcommand{\questionMostMemorablePart}{\enquote{What was the most memorable part of the game for you?} [open]}
\newcommand{\qFavoriteBoardGame}{\enquote{Favorite Board Game?} [open]}
\newcommand{\questionFavoriteBoardGame}{\enquote{Which board game do you enjoy playing the most?} [open]}
\newcommand{\qFrequencyOfSituations}{\enquote{Frequency of Situations?}}
\newcommand{\questionFrequencyOfSituations}{\enquote{How often do software developers find themselves in similar situations?}}
\newcommand{\qGameSImpactOnDev}{\enquote{Game's Impact on Dev?}}
\newcommand{\questionGameSImpactOnDev}{\enquote{To what extent did the game experience contribute to your professional development in software development?}}
\newcommand{\qOpinionOnOwnCommitClarity}{\enquote{Opinion on Own Commit Clarity?}}
\newcommand{\questionOpinionOnOwnCommitClarity}{\enquote{What is your opinion on the clarity of your own commit messages?}}
\newcommand{\qDullestMoment}{\enquote{Dullest Moment?} [open]}
\newcommand{\questionDullestMoment}{\enquote{When was the game the most boring or uncomfortable for you?} [open]}
\newcommand{\qRolesToday}{\enquote{Roles Today?}}
\newcommand{\questionRolesToday}{\enquote{What roles did you play today?}}
\newcommand{\qPlayedRoles}{\enquote{Played Roles?}}
\newcommand{\questionPlayedRoles}{\enquote{What roles did you play?}}
\newcommand{\qBoardGameExperience}{\enquote{Board Game Experience?}}
\newcommand{\questionBoardGameExperience}{\enquote{What are your experiences with board games in general?}}
\newcommand{\qEnhanceGame}{\enquote{Enhance Game?} [open]}
\newcommand{\questionEnhanceGame}{\enquote{How would you complement the game? How would you make it even more enjoyable?} [open]}
\newcommand{\qJustifyFutureCommitChoices}{\enquote{Justify Future Commit Choices} [open]}
\newcommand{\questionJustifyFutureCommitChoices}{\enquote{Why did you make these decisions regarding future commit messages?} [open]}
\newcommand{\qProjectChoiceReasons}{\enquote{Project Choice Reasons} [open]}
\newcommand{\questionProjectChoiceReasons}{\enquote{Why did you choose the projects provided during registration? Were there any special circumstances in your choice?} [open]}
\newcommand{\qChangesInCommitsWriting}{\enquote{Changes in Commits' Writing?}}
\newcommand{\questionChangesInCommitsWriting}{\enquote{Will you write commit messages differently after the game?}}
\newcommand{\qSurprisingResponses}{\enquote{Surprising Responses?}}
\newcommand{\questionSurprisingResponses}{\enquote{Did the responses of colleagues and authors to commit messages surprise you?}}
\newcommand{\qSurprisingResponsesToIssues}{\enquote{Surprising Responses to Issues?}}
\newcommand{\qMainWorkArea}{\enquote{Main Work Area?} [open]}
\newcommand{\questionMainWorkArea}{\enquote{Briefly describe the area in which you work or have worked the most! (programming languages, technologies, methodologies, etc.)} [open]}
\newcommand{\qGameSPurpose}{\enquote{Game's Purpose?} [open]}
\newcommand{\questionGameSPurpose}{\enquote{In your own words, explain what you think the purpose of the game was!} [open]}
\newcommand{\qGameSSuccess}{\enquote{Game's Success?}}
\newcommand{\questionGameSSuccess}{\enquote{Do you think the game achieved its goal?}}
\newcommand{\qGameImpactOutsideWork}{\enquote{Game Impact Outside Work?} [open]}
\newcommand{\questionGameImpactOutsideWork}{\enquote{Do you think the experiences gained during the game will have an impact on your non-professional life? If yes, please specify!} [open]}
\newcommand{\qWhyGameSucceededFailed}{\enquote{Why Game Succeeded/Failed?} [open]}
\newcommand{\questionWhyGameSucceededFailed}{\enquote{In your opinion, why did the game succeed or fail in achieving its goal?} [open]}
\newcommand{\qPositiveParticipation}{\enquote{Positive Participation?}}
\newcommand{\questionPositiveParticipation}{\enquote{Was it positive for you to participate in the game? Did you enjoy the game?}}
\newcommand{\qAdditionalComments}{\enquote{Additional Comments?} [open]}
\newcommand{\questionAdditionalComments}{\enquote{Do you have any other comments or suggestions regarding the game?} [open]}
\newcommand{\qSimilarSituationsBefore}{\enquote{Similar Situations Before?}}
\newcommand{\questionSimilarSituationsBefore}{\enquote{Have you been in a similar situation before? If yes, in what role?}}